\documentclass{aastex}
\usepackage{emulateapj5}

\usepackage{epsfig}

\def \wdvd{WD$\,$2211$-$495}
\def \ie{{\it i.e.}}
\def \eg{{\it e.g.}}
\def \kid{$\chi^2$}
\def \deltakid{$\Delta\chi^2$}

\def \lya{Lyman~$\alpha$}

\def \cmmd{cm$^{-2}$}
\def \kms{${\rm km\,s}^{-1}$}
\def \dshism{(D/H)$_{\mathrm{ISM}}$}

\def \ow{{\tt Owens.f}}
\def \ism{interstellar medium}
\def \fpn{fixed-pattern noise}
\def \lsf{line spread function}

\def \fuse{{\it FUSE}}
\def \ds{$2\,\sigma$}

\shorttitle{Deuterium Abundance Toward WD$\,$2211$-$495$\,$: 
Results from the {\it FUSE} Mission}
\shortauthors{H\'ebrard et al.}

\begin{document}

\title{Deuterium Abundance Toward WD$\,$2211$-$495$\,$: 
Results from the Far Ultraviolet Spectroscopic 
Explorer ({\it FUSE}) Mission
}


\author{G. H\'ebrard\altaffilmark{2}, 
M. Lemoine, A. Vidal-Madjar, J.-M. D\'esert, \\
A. Lecavelier des \'Etangs, R.~Ferlet}
\affil{Institut d'Astrophysique de Paris, 98$^{bis}$ boulevard Arago, 
        F-75014 Paris, France}

\author{B. E. Wood, J. L. Linsky}
\affil{JILA, University of Colorado and NIST, Campus Box 440, Boulder, 
        CO 80309-0440, USA}

\author{J. W. Kruk, P. Chayer\altaffilmark{3}, S. Lacour, W. P. Blair, \\
S. D. Friedman, H. W. Moos, K. R.~Sembach}
\affil{Department of Physics and Astronomy, Johns Hopkins University, 
         Baltimore, MD 21218, USA}

\author{G. Sonneborn, W. R. Oegerle}
\affil{Laboratory for Astronomy and Solar Physics, NASA/GSFC, Code 681, 
         Greenbelt, MD 20771, USA}

\and

\author{E. B. Jenkins}
\affil{Princeton University Observatory, Princeton, NJ 08544, USA}

\altaffiltext{2}{email: hebrard@iap.fr}

\altaffiltext{3}{Primary affiliation: Department of Physics and Astronomy, 
  University of Victoria, P.O. Box 3055, Victoria, BC V8W 3P6, Canada. }

\begin{abstract}
We present a deuterium abundance analysis of the line of sight 
toward the white dwarf \wdvd\ observed with the Far Ultraviolet 
Spectroscopic Explorer (\fuse). Numerous interstellar lines are 
detected on the continuum of the stellar spectrum. 
A~thorough analysis was performed through the simultaneous fit of 
interstellar absorption lines detected in the four \fuse\ channels 
of multiple observations with different slits. 
We excluded all saturated lines in order to reduce possible systematic 
errors on the column density~measurements. 
We report the determination of the average interstellar D/O 
and D/N ratios along this line of~sight at the 95\% 
confidence level: 
D/O$\;= 4.0 \, (\pm1.2) \times 10^{-2}$;
D/N$\;= 4.4 \, (\pm1.3) \times 10^{-1}$. 
In conjunction with \fuse\ observations of other nearby sight lines,
the results of this study will allow a deeper understanding of the 
present-day abundance of deuterium in the local interstellar medium
and its evolution with time.
	\end{abstract}


\keywords{ISM: abundances -- ISM: clouds -- cosmology: observations -- 
ultraviolet: ISM -- stars: individual (WD2211-495) -- (stars:) white dwarfs}


\section{Introduction}

Deuterium is believed to be produced in appreciable quantities 
only in primordial Big Bang nucleosynthesis (BBN) and destroyed in stellar 
interiors (\eg, Epstein et al.~\citealp{epstein76}); it is thus a 
key element in cosmology and in galactic chemical evolution
(see \eg~Vangioni-Flam \& Cass\'e~\citealp{flam95}; 
Prantzos~\citealp{prantzos96}; Scully et al.~\citealp{scully97}). 
The primordial abundance of deuterium is one of the best probes of 
$\Omega_Bh^2$, the baryonic density of the Universe divided by the 
critical density. 
The abundance of deuterium relative to hydrogen 
(D/H) is expected to decline during 
Galactic evolution at a rate that is a function of the star formation 
rate; standard models predict a factor of 2 to 3 decrease in the
deuterium abundance in 15 Gyrs (see \eg, Galli et al.~\citealp{galli95}; 
Tosi et al.~\citealp{tosi98}). Hence, any
abundance of deuterium measured at any metallicity should provide a
lower limit to the primordial deuterium abundance. 
This picture is essentially constrained by deuterium abundance 
measurements at look-back times of $\sim14$~Gyrs 
(primordial intergalactic clouds), 
4.5~Gyrs (protosolar), and 0.0~Gyrs (\ism). 
Although the evolution of the deuterium abundance seems to be 
qualitatively understood, measurements of D/H at similar redshifts 
show some dispersion and indicate that additional processes may be 
important in controlling the abundance of deuterium. 
That fact has led to the development of non-standard models, which 
propose, for example, larger astration factors 
(\eg, Vangioni-Flam et al.~\citealp{flam94}) 
or non-primordial deuterium production 
(see~\eg, Lemoine et al.~(\citealp{lemoine99}) for a~review).

Up to now, the \ism\ is the astrophysical site that has allowed the most 
comprehensive investigations of deuterium abundances. 
Deuterium has been observed in the \ism\ using different methods: 
radio measurements of its 92~cm hyperfine transition
(\eg, Blitz \& Heiles~\citealp{blitz87}; 
Chengalur et al.~\citealp{chenga97}), observations of 
deuterated molecules (\eg, Lubowich et al.~\citealp{lubo00}; 
Ferlet et al.~\citealp{ferlet00}), Balmer series analyses 
(H\'ebrard et al.~\citealp{hebrard00}), and Lyman series
absorption (see Moos et al.~\citealp{moos01}). 
Of these, the most accurate measurements have been obtained through 
Lyman absorption-line observations in the far-ultraviolet (far-UV) 
spectral range. 
By observing hydrogen and deuterium directly in their atomic form,
far-UV Lyman series absorption-line measurements provide accurate
column density determinations that are not dependent on ionization 
or chemical fractionation effects.
The first measurement of the abundance ratio \dshism\ was reported 
by Rogerson \& York~(\citealp{ry73}) for the line of sight to 
$\beta$~Cen, using {\it Copernicus}: \dshism$=1.4\pm0.2\times10^{-5}$. 
Since then, many other \dshism\ measurements have been performed using 
different instruments ({\it Copernicus}, {\it IUE}, {\it IMAPS}, 
{\it HST}) for other 
sight lines, and the values obtained show significant dispersion 
around the above~value. 

For example, an average value \dshism$=1.50(\pm0.10)\times10^{-5}$ 
($1\,\sigma$) has been 
derived for the Local Interstellar Cloud 
(Lallement \& Bertin~\citealp{lallement92}) 
by Linsky~(\citealp{linsky98}) from the comparison of 12 nearby 
sight lines, but studies of several lines of sight revealed 
values outside this range (\eg,~Laurent et al.~\citealp{laurent79}; 
York~\citealp{york83}; Allen et al.~\citealp{allen92}; 
Vidal-Madjar et al.~\citealp{avm98}; H\'ebrard et 
al.~\citealp{hebrard99}; Jenkins et al.~\citealp{jenkins99}; 
Sonneborn et al.~\citealp{sonneborn00}). 
This dispersion may result from spatial variations 
due to some unknown physical processes or 
underestimation of systematic errors. 
There is still considerable debate over these two interpretations,
and the final resolution of the issue may have implications for 
understanding the physics of the interstellar medium, as well as 
the baryonic density inferred from D/H measurements.
Measurements of D/H in the intergalactic medium are also sparse and
do not agree with each other (\eg,~Webb et al.~\citealp{webb97}; 
Burles \& Tytler~\citealp{burles98a}; \citealp{burles98b}), 
so an accurate determination of the primordial 
deuterium abundance has also proven to be a difficult quantity 
to measure.  Moreover, recent studies of the anisotropy of the 
cosmic microwave background (CMB), which permits evaluations of the 
baryonic density independent of those obtained through deuterium 
measurements (see, \eg, de~Bernardis et al.~\citealp{bernardis00}; 
Jaffe et al.~\citealp{jaffe01}), imply higher values of $\Omega_Bh^2$ 
than those implied by the abundance studies, corresponding to a 
primordial D/H value of the order of the \dshism~values 
(see however de~Bernardis et al.~\citealp{bernardis01}; 
Stompor et al.~\citealp{stompor01}; Pryke et al.~\citealp{pryke01}).

An accurate determination of the interstellar deuterium abundance
is one of the main objectives of the 
Far Ultraviolet Spectroscopic Explorer (\fuse), which was 
successfully launched on 1999 June 24 (Moos et al.~\citealp{moos00}). 
In this paper, 
we present new measurements of deuterium abundances obtained with 
\fuse\ toward the white dwarf \wdvd. 
This paper is part of a series dedicated 
to \fuse\ measurements of deuterium interstellar abundances toward  
BD\,+28$^{\circ}$\,4211 (Sonneborn et al.~\citealp{sonneborn01}), 
WD$\,$1634$-$573 (Wood et al.~\citealp{wood01}), 
WD$\,$0621$-$376 (Lehner et al.~\citealp{lehner01}), 
G191$-$B2B (Lemoine et al.~\citealp{lemoine01}), 
HZ$\,$43A (Kruk et al.~\citealp{kruk01}), and Feige~110 
(Friedman et al.~\citealp{friedman01}). 
A major goal of this program is to determine to what extent and on 
what scales variations in the D/H ratio occur. 
Moos et al.~(\citealp{moos01}) present an overview of 
these first results. 

White dwarfs are ideal targets for the 
measurement of the interstellar deuterium abundance (Lemoine et 
al.~\citealp{lemoine96}); they may be chosen close to the
Sun (so that the column densities are not too high and the
velocity structure of the line of sight is not too complex), and 
they exhibit relatively smooth UV continua (which allow detections 
of lines from many species). 
Prior to this series of papers, measurements have been 
published for \dshism\ toward three 
white dwarfs: G191$-$B2B (Lemoine et al.~\citealp{lemoine96}; 
Vidal-Madjar et al.~\citealp{avm98}; Sahu et al.~\citealp{sahu99}), 
Sirius-B (H\'ebrard et al.~\citealp{hebrard99}), and 
Feige~24 (Vennes et al.~\citealp{vennes00}).

The \fuse\ observations of \wdvd\ and the data processing are presented 
in Sect.~\ref{Observations_and_data_processing}, and the details of 
the analysis in Sect.~\ref{Data_Analysis}. The results are reported 
in Sect.~\ref{Results} and discussed in Sect.~\ref{Discussion}.

\section{Observations and data processing}
\label{Observations_and_data_processing}

\wdvd\ (RE~J2214$-$491) was identified by the {\it ROSAT} wide
field camera all-sky survey as an extreme ultraviolet source and
was classified as a new white dwarf star 
(Pounds et al.~\citealp{pounds93}). Holberg et al.~(\citealp{holberg93})
analyzed the optical and high-resolution 
International Ultraviolet Explorer ({\it IUE}) spectra
of \wdvd\ and concluded it was a hot DA white dwarf showing
traces of heavy elements such as C, N, O, Si, and Fe.  Later, 
Holberg et al.~(\citealp{holberg94}) and Werner \& 
Dreizler~(\citealp{werner94}) discovered the presence of
Ni in the co-added {\it IUE} spectrum, confirming that WD~2211$-$495
was one of the most metal-rich white dwarfs known.  Barstow et al.\ (1998)
determined the atmospheric parameters of the star by considering
non-LTE metal-line-blanketed model atmospheres and found
$T_{\rm{eff}} = 62$,000 K and $\log g = 7.2$.  WD~2211$-$495 has a
mass $M \simeq 0.53$ M$_\sun$ according to optical spectroscopic
analyses performed by Marsh et al.~(\citealp{marsh97a}), 
Vennes et al.~(\citealp{vennes97}), and Finley et al.~(\citealp{finley97}). 
Its photometric distance is $d\simeq53$pc 
(Holberg et al.~\citealp{holberg98}). 
Table~\ref{wd_parameters} summarizes relevant sight line and 
atmospheric parameters for \wdvd.

\subsection{Observations}

\wdvd\ was observed in 1999 and 2000 as part of the \fuse\ photometric 
calibration program and the \fuse\ Science Team D/H program. 
Seventeen observations consisting of 188 individual exposures for 
a total exposure time of $93\,800$~sec ($\sim26$~hours) were used in 
the present study. 
All the observations were obtained in histogram mode, through one 
of the three slits available (Moos et al.~\citealp{moos00}): the 
large (LWRS), medium (MDRS), or high resolution aperture (HIRS). 
Table~\ref{table_obs} shows the log of the observations used 
in this analysis.

\subsection{Data processing}
\label{Data_processing}

The one-dimensional spectra were extracted from the two-dimensional 
detector images and calibrated using slightly different versions 
of the {\tt CALFUSE} pipeline (see Table~\ref{table_obs}), 
depending of the version available in the archives for a given 
observation. 
For a given slit (LWRS, MDRS, or HIRS), data from each channel 
and segment (SiC1A, SiC2B, LiF1A, LiF2B, etc.) were co-added 
separately, after 
wavelength shift corrections of the individual calibrated exposures. 
Wavelength shifts between individual spectra were typically a few 
pixels for exposures of a given observation and a few tens of pixels 
between exposures of different observations. 
Because a given wavelength falls on different pixels in different 
exposures, co-addition of the numerous exposures results in a 
considerable reduction in \fpn\ (Sahnow et al.~\citealp{sahnow00}) 
by acting like a random FP-Split procedure (Kruk et al.~\citealp{kruk01}). 

Exposures with strong airglow emission, large spectral resolution 
variations, or large wavelength stretching\footnote{Wavelength stretchings 
are inconsistencies in the wavelength scale 
(Sahnow et al.~\citealp{sahnow00}).} were not included in the sums. 
The last two criteria were applied in particular to the 
two observations performed in 1999, early in the mission. 
In order to properly co-add different exposures for a given segment, 
the same wavelength calibration file ({\tt wave*009}) was used in 
{\tt CALFUSE} for all of the observations. That led us to re-run 
{\tt CALFUSE} in the version 1.8.7 (which includes {\tt wave*009}) 
for the observations which were archived in an older version which 
did not include this calibration file (see Table~\ref{table_obs}). 
The spectral resolution in the final spectra ranges between $\sim13000$ 
and $\sim18500$, depending on detector segment and wavelength 
(see Section~\ref{parameters_values}).  

\begin{figure}
\psfig{file=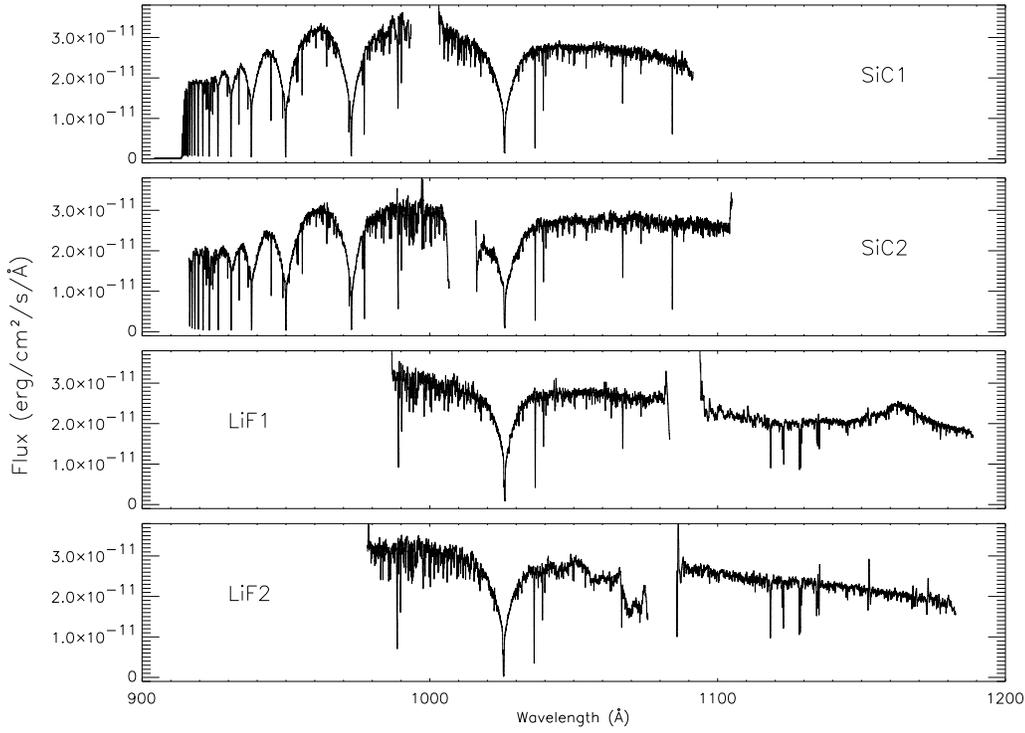,height=10cm}
\caption{The eight segments of the \fuse\ \wdvd\ spectra obtained 
through the large aperture (LWRS). 
Each channel (SiC1, SiC2, LiF1, and LiF2) is divided in two segments, 
separated by a gap. 
Similiar spectra were obtained for the two other slits (MDRS and HIRS); 
in total, $8\times3=24$~co-added spectra were used for the analysis. 
The bump in the LiF1 continuum at 1150\AA\ is an instrumental artifact. 
\label{fig_spectre_all}}
\end{figure}

\begin{figure}
\psfig{file=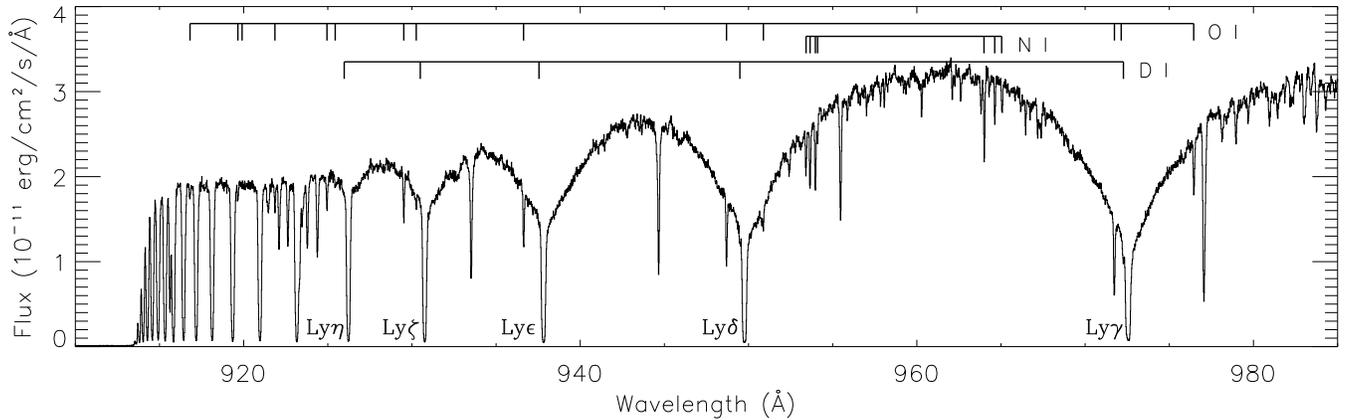,height=6cm}
\caption{\fuse\ SiC1B spectrum of \wdvd\ observed through the LWRS
slit. Positions 
of \ion{D}{1}, \ion{O}{1}, and \ion{N}{1} interstellar absorption 
lines are~indicated. 
\label{fig_spectre_lines}}
\end{figure}

All of the spectra were binned to three pixel samples (the \lsf\ (LSF) 
is about 10~pixels wide). 
The final LWRS spectra are plotted in Figure~\ref{fig_spectre_all}; 
similar datasets were obtained for the MDRS and HIRS slits.
A sample of a spectrum (SiC1B with LWRS) is shown 
in Figure~\ref{fig_spectre_lines} with the \ion{D}{1}, \ion{O}{1}, 
and \ion{N}{1} lines identified.

\section{Data Analysis}
\label{Data_Analysis}

\subsection{Method overview}
\label{method}

For a given transition of a given element, each interstellar cloud 
along the sight line produces an absorption line that can be 
modeled with a Voigt profile. In addition to atomic parameters 
[taken from Morton~(\citealp{morton91,morton99})] a Voigt profile 
is defined by four cloud 
parameters: the radial velocity $v$ of the cloud (in \kms), the column 
density $N_i$ of the element $i$ (in cm$^{-2}$), the temperature $T$ 
of the gas (in~K), and its micro-turbulent velocity $\xi$ (in~\kms). 
\fuse\ does not have sufficient spectral resolution 
to resolve the individual velocity components along the 
\wdvd\ sight line. This implies that \fuse\ cannot resolve the 
fine shape of individual velocity components and that it
cannot reveal the presence of closely spaced multiple components. 
We do not have high-resolution ground-based interstellar 
data at our disposal for this star. 

The analysis has been done using the profile fitting procedure
\ow, developed by one of us (M.L.) and the \fuse\ French Team. 
A subset of the data has been independently fitted using a second 
procedure (see Sect.~\ref{discussion_of_the_effects_and_tests}), 
developed by another of us (B.E.W.). The reader should refer to 
Wood et al.~(\citealp{wood01}) for a full description of this 
fitting procedure, and we will concentrate on describing in 
detail the results obtained with \ow. 

\begin{figure}
\psfig{file=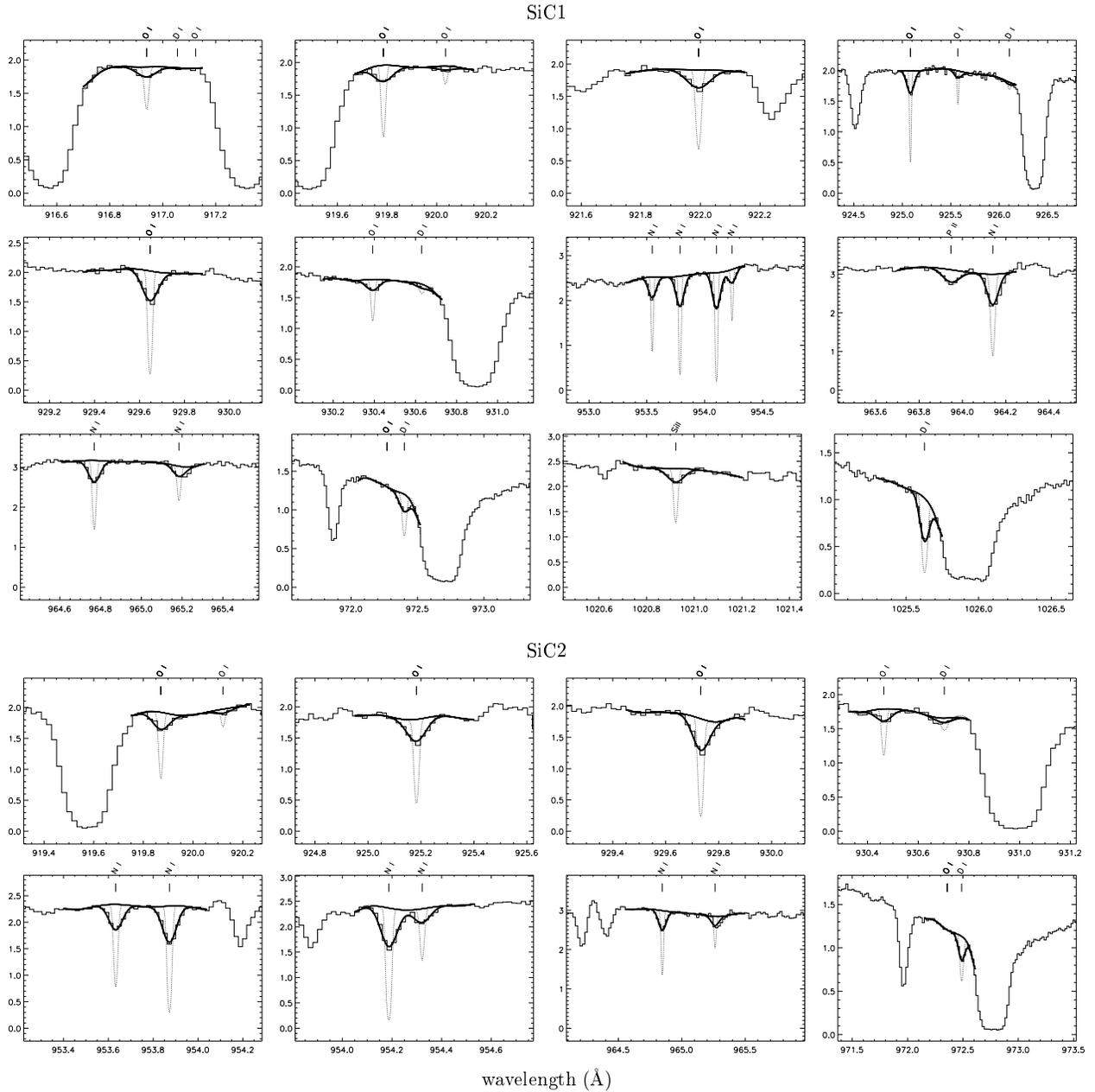,height=25.5cm}
\vspace{-6cm}
\caption{
Plots of the fits to lines from the LWRS SiC data.  The upper 12 spectra
windows are from the SiC1 channel and the lower 8 spectral windows are
from SiC2. Histogram lines are the data, the solid thick lines are the 
fits and continua, and the dotted lines are the model profiles prior to 
convolution with the LSF. 
Y-axis is flux in $10^{-11}\,$erg/cm$^2$/s/\AA.
\label{fig_fit_L}}
\end{figure}

\begin{figure}
\psfig{file=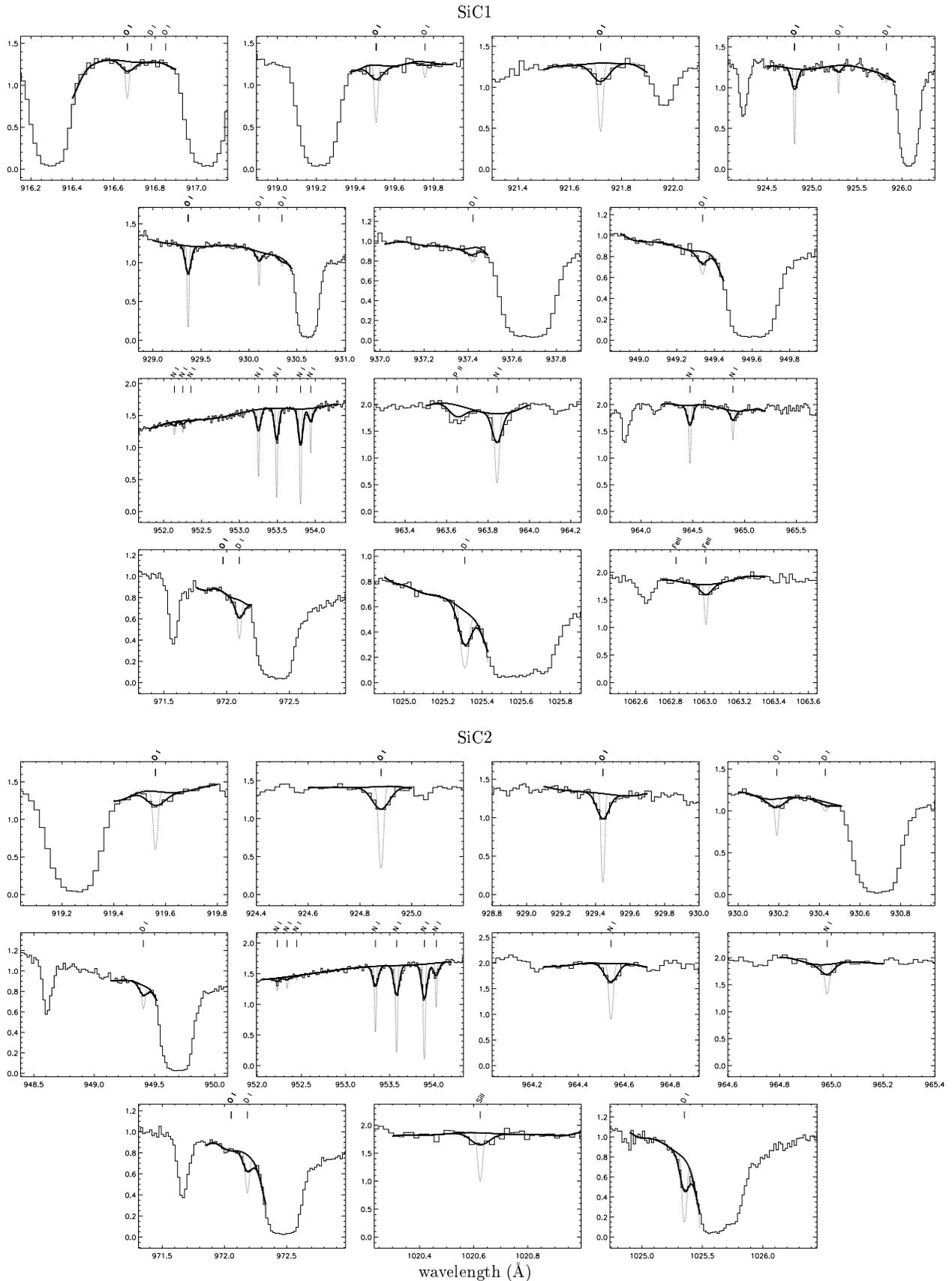,height=25.5cm}
\vspace{-1.3cm}
\caption{Same as Fig.~\ref{fig_fit_L}, for MDRS SiC1 (upper) and 
SiC2 (lower) spectral windows.\label{fig_fit_Ms}}
\end{figure}

\begin{figure}
\psfig{file=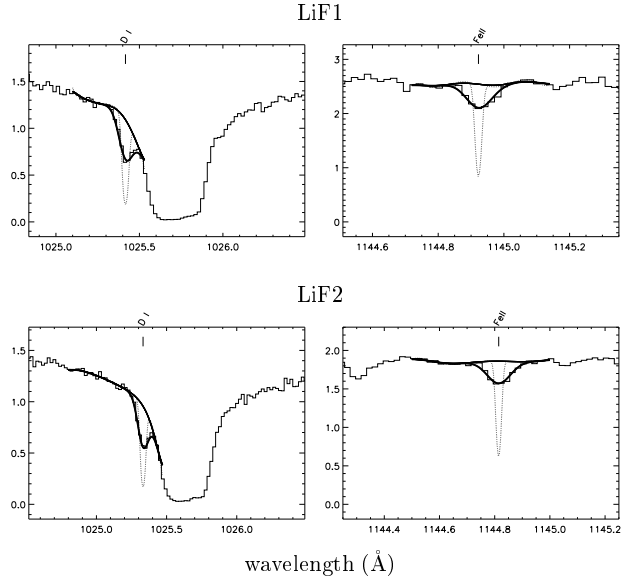,height=25.5cm}
\vspace{-15cm}
\caption{Same as Fig.~\ref{fig_fit_L}, for MDRS LiF1 (upper) and 
LiF2 (lower) spectral windows. \label{fig_fit_Ml}}
\end{figure}

\begin{figure}
\psfig{file=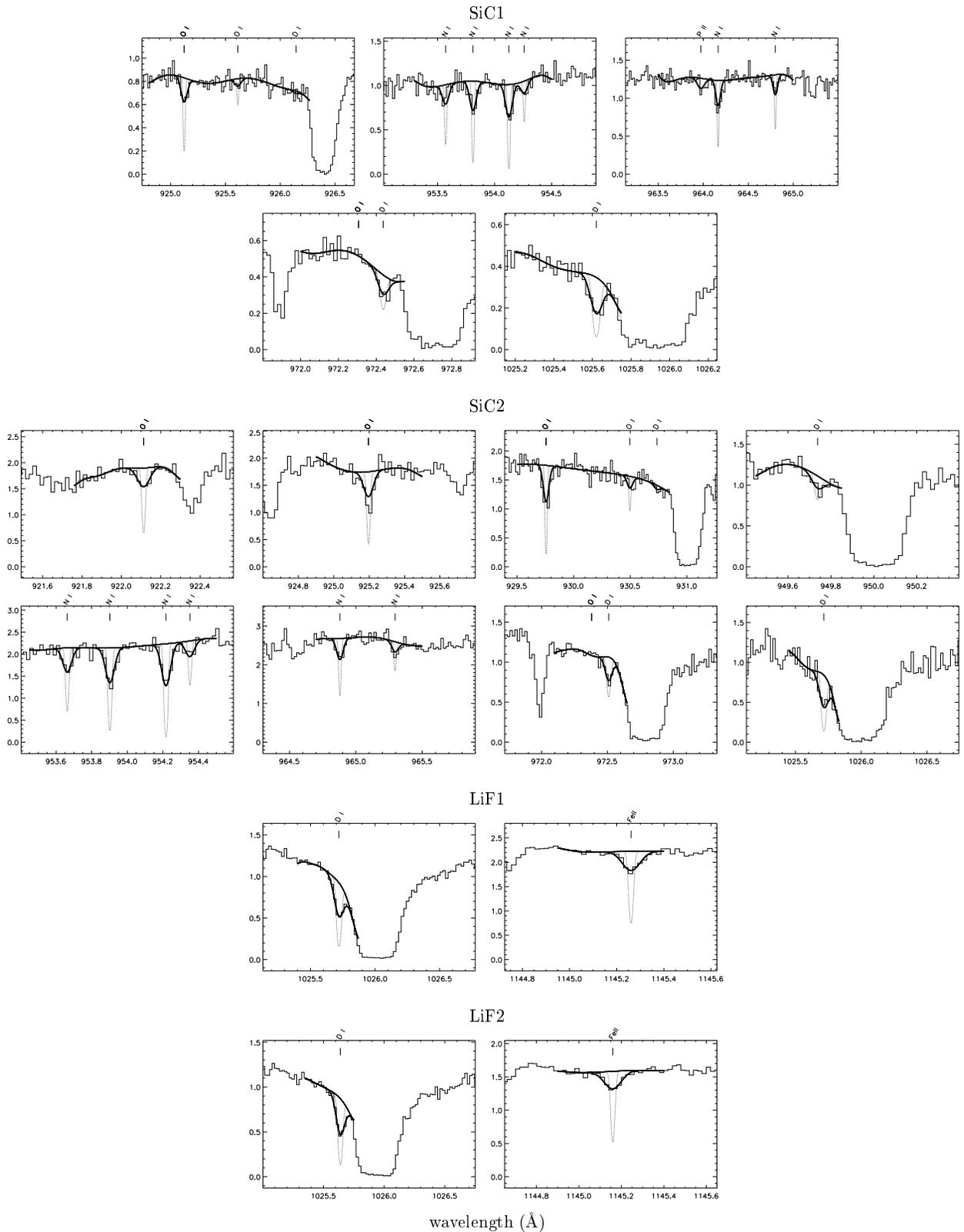,height=25.5cm}
\vspace{-2.7cm}
\caption{Same as Fig.~\ref{fig_fit_L}, for HIRS SiC1, SiC2, 
LiF1 and LiF2 spectral windows. \label{fig_fit_H}}
\end{figure}

\ow\ models Voigt profiles using \kid\ minimization. 
One of the characteristics of this procedure is that it is able to fit 
simultaneously several lines in different spectral windows, 
a spectral window being a part of a full spectrum 
(see Figures~\ref{fig_fit_L}-\ref{fig_fit_H}). 
Thus, this procedure finds the best solution compatible with 
numerous spectral lines based on the 
assumption that all lines considered yield the same values for $v$, $T$,
and $\xi$ for a given component, and that all lines for a specific 
component of element $i$ give the same value for $N_i$. 
The width of the lines 
combines both parameters $T$ and $\xi$, 
which can only be separately determined if several elements with 
different masses are simultaneously fit\footnote{The $b$-value, 
which is the width of a line, is given by $b^2 = \frac{2kT}{m} + {\xi}^2$, 
where $k$ is the Boltzmann constant and $m$ the mass of the element.}. 

Data at the edge of each detector segment are inaccurate 
(see Figure~\ref{fig_spectre_all}) and were not used in the fit. 
In the end, 32 different absorption lines (\ion{D}{1}, \ion{O}{1}, 
\ion{N}{1}, \ion{Fe}{2}, \ion{Si}{2}, and \ion{P}{2}) were modeled, 
which were observed in different segments and spectra, resulting
in a total of 115 simultaneously fitted lines. 
Remaining ions like \ion{Ar}{1} and 
\ion{N}{2} were modeled separately in extra fits. 
No H$_2$ lines were detected. 
The 115 fitted lines are located in 65 different 
spectral windows, extracted from the different spectra, obtained 
through the HIRS, MDRS, and LWRS slits, on channels SiC1, SiC2, LiF1, 
and LiF2. Plots of the 65 spectral windows are shown in 
Figures~\ref{fig_fit_L}-\ref{fig_fit_H}. 
Atomic parameters of the \ion{D}{1}, \ion{O}{1}, 
and \ion{N}{1} lines used are reported in~Table~\ref{table_lines}.

\subsection{Free parameters}
\label{free_parameters}

All the cloud parameters ($v$, $T$, $\xi$ and all the $N_i$'s)
were free to vary in the fit. 
We also allowed the continua shape, the spectral shift and the LSF 
width to vary from one spectral window to the next. These 
parameters are a priori unknown or poorly known, and this uncertainty 
is likely to introduce systematic effects in the final result. 
We take into account this uncertainty by treating these unknowns 
as free parameters in the fit, so that their uncertainty is included 
in the \kid\ variations (see Sect.~\ref{chi2_scaling}).

The continua were fit by $3^{rd}$ to $7^{th}$ order polynomials
(see Figures~\ref{fig_fit_L}-\ref{fig_fit_H}), depending on the
spectral region. 
All the parameters of the polynomials were free. 
The spectral shift of each window\footnote{Actually, the spectral shift 
was free for all windows but one in order to avoid degeneracy in a 
parameter space with the radial velocity of the absorber.} 
(\ie, its wavelength zero point) 
was free to allow for possible inaccuracy of the \fuse\ 
wavelength calibration. 

All Voigt profiles were convolved with a simple Gaussian LSF. 
The spectral resolution of \fuse\ is not accurately determined, but 
it is known to depend on the segment and wavelength. 
We let the width of the LSF vary freely for each spectral window. 
Note that fits with a double Gaussian LSF were also performed 
(see Sect.~\ref{discussion_of_the_effects_and_tests}).

\subsection{Unsaturated lines}

Since we are primarily interested in 
abundances, we focused on the determination of column densities. 
We chose to use only unsaturated lines in the fits to reduce 
uncertainties resulting from saturation effects and other systematic
errors. Saturated lines are not very sensitive to the column density. 
Since the lines which we use are on the linear part of the curve of 
growth, we can measure the total column densities for the line of sight 
even though the individual clouds are not resolved. 
All the \ion{H}{1} interstellar lines detectable in the \fuse\ spectral
range are located on the flat part of the curve of growth for this 
sight line; hence, no accurate \ion{H}{1} column density could be 
determined. 
We thus determined D/O and D/N ratios instead of D/H. 
In order to fit the Lyman \ion{D}{1} lines without fitting \ion{H}{1} 
lines with Voigt profiles, the blue wings of the 
\ion{H}{1} lines were fit with polynomials 
(see \eg, Figure~\ref{fig_fit_Ml}, 
left panel) to provide the continuum for \ion{D}{1} absorption
(see, however, Sect.~\ref{discussion_of_the_effects_and_tests}).

\subsection{Stellar model}
\label{stellar_model}

We compared the \fuse\ spectrum of \wdvd\ to a stellar
model in order to verify that the interstellar lines were not blended 
with stellar lines.  To this end, we computed a non-LTE metal
line-blanketed model atmosphere using the programs {\tt TLUSTY/SYNSPEC} 
(Hubeny \& Lanz~\citealp{hubeny95}).  
We used the atmospheric parameters determined
by Barstow et al.~(\citealp{barstow98}) (see Table~\ref{wd_parameters}).  
We performed an abundance
analysis using the \fuse\ and {\it IUE} data and then computed a
synthetic spectrum incorporating these abundances 
(Chayer et al.~\citealp{chayer01}). 
The comparison between the stellar model and
the \fuse\ spectrum allowed us to exclude the \ion{O}{1}
$\lambda$950.88\AA\ and \ion{Si}{2} $\lambda$989.87\AA\ interstellar 
lines that are blended with the \ion{P}{4} $\lambda$950.66\AA\ and
\ion{N}{3} $\lambda$989.80\AA\ stellar lines.
In agreement with the DA-type classification of the \wdvd, 
no \ion{He}{2} absorption is seen in the \fuse\ spectrum 
(\eg, $\lambda1084.9$\AA\ or $\lambda992.3$\AA), and hence such absorption 
should be absent at the intervening \ion{H}{1} Lyman lines.

\subsection{Zero flux level}
\label{zero_flux_level}

To properly deduce abundances from interstellar absorption lines, it 
is extremely important to precisely evaluate the zero flux level. 
Although scattered light in \fuse\ is low 
(Moos et al.~\citealp{moos00}; Sahnow et al.~\citealp{sahnow00}), 
some residual flux is visible on the bottom of the \ion{H}{1} Lyman 
lines, as one can see in Figures~\ref{fig_spectre_all} 
and~\ref{fig_spectre_lines}. 
This residual flux (at a level of few percent of the continuum) 
is probably due to the scatter of adjacent flux from the continuum 
due to the wings of the LSF. 
This interpretation agrees with the fact that this 
small amount of flux is not detected at wavelengths lower than the 
Lyman break, where 
there is no flux for scatter. Since the residual flux is so low, for 
each spectral window we tuned the zero flux level to the signal at 
the core of the nearest Lyman line, where it is
readily measurable because these 
lines are totally saturated. The diffuse light is a few percent 
of the continuum, and its variations as a function of wavelength 
seem to be of the same order of magnitude, or even lower. 
Fits with a double Gaussian LSF were also performed 
(see Sect.~\ref{discussion_of_the_effects_and_tests}).

\subsection{\kid\ scaling and \deltakid\ method}
\label{chi2_scaling}

The \kid\ of the final fit was 2424.2. 
The difference between the number of independent spectral bins
used in the fit and the number of free parameters (cloud, polynomials, 
shifts, and LSF parameters) was 1764. For this number of degrees of 
freedom, the reduced \kid\ is $1.37$. 
This value is larger than $1.00$, probably because of a small 
underestimation of the tabulated errors for each spectral bin.
These errors are computed by the extraction 
pipeline which does not take into account all possible instrumental 
artifacts. 

\begin{figure}
\psfig{file=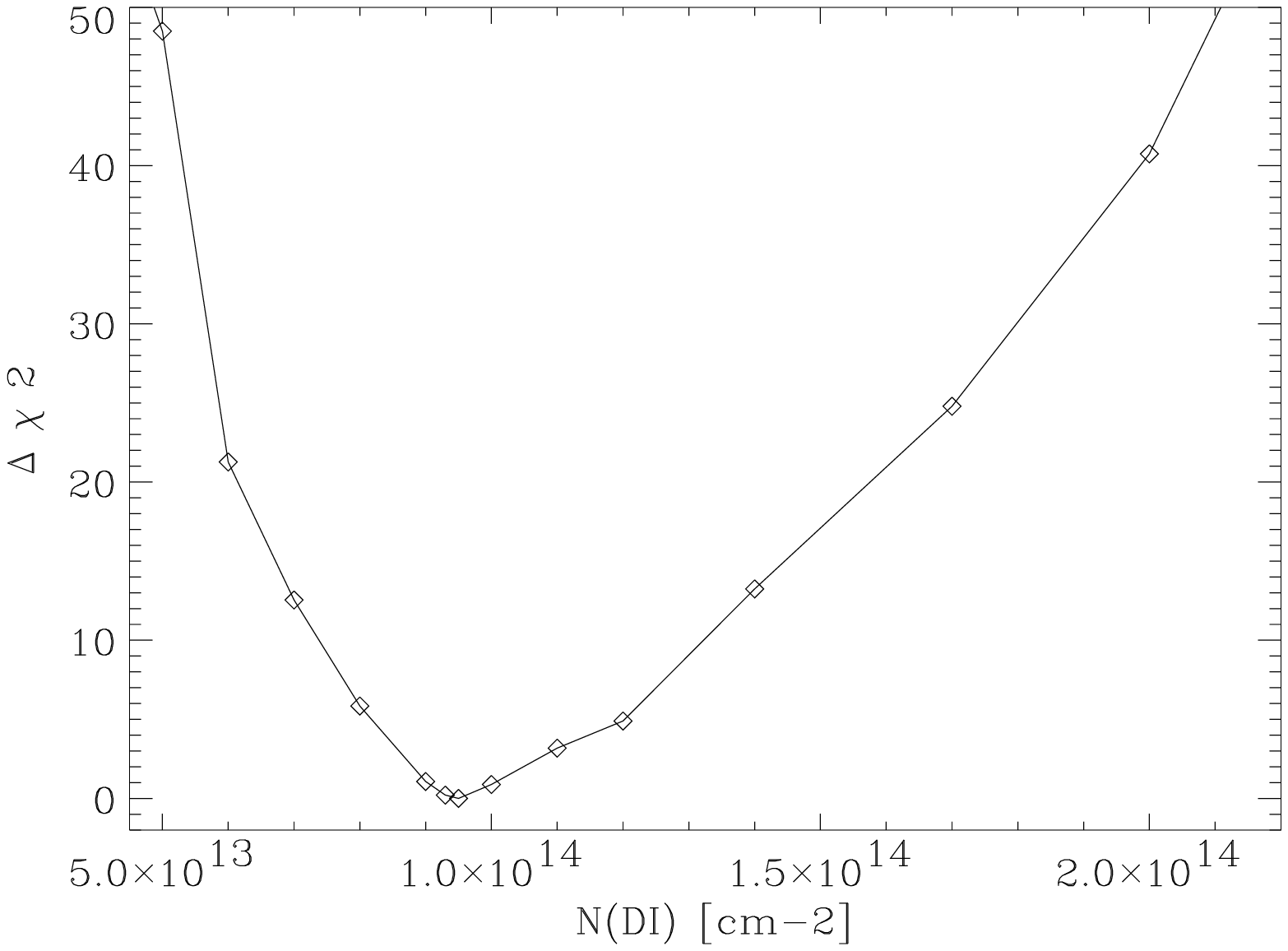,height=4.5cm}
\psfig{file=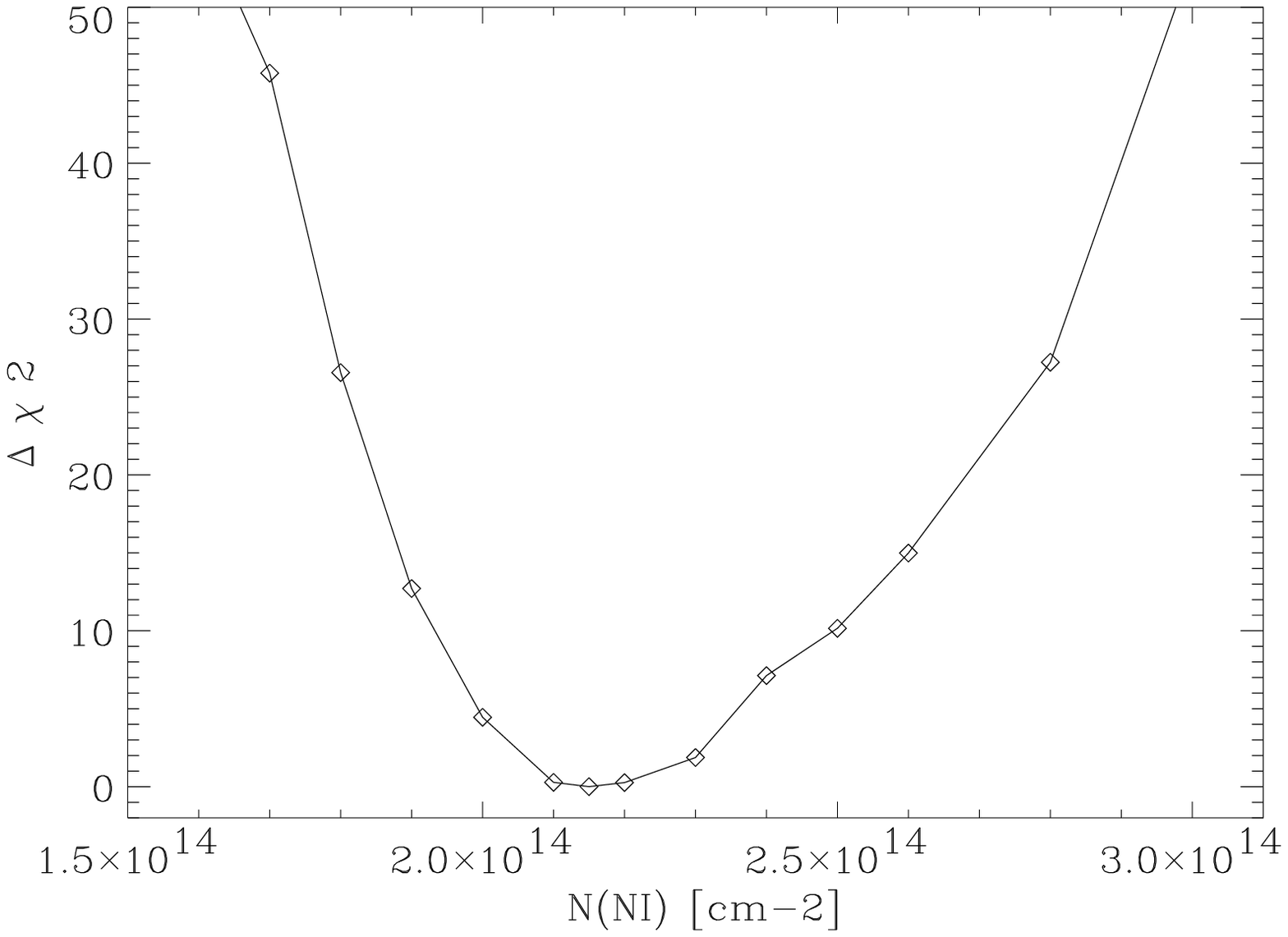,height=4.5cm}
\psfig{file=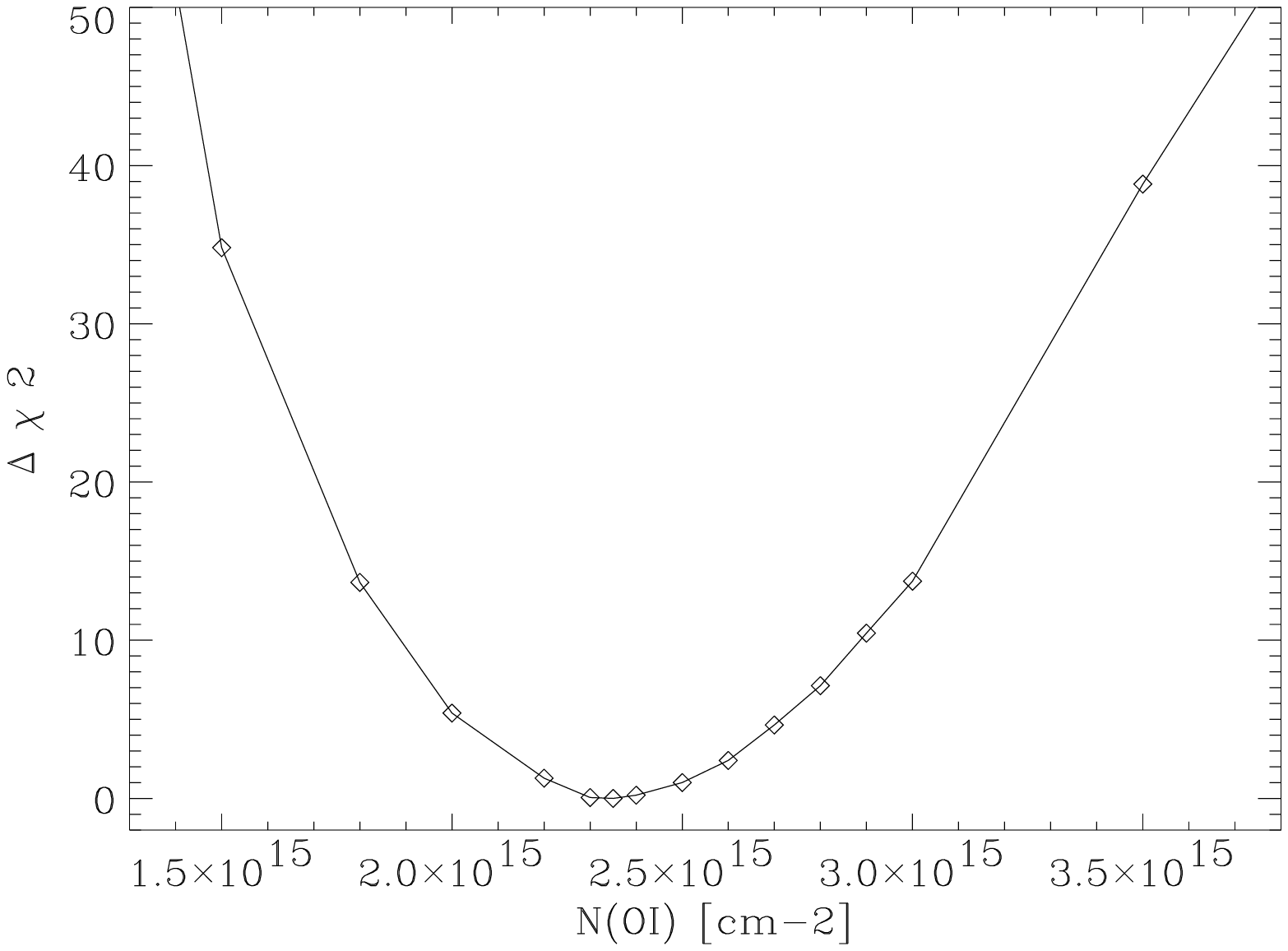,height=4.5cm}
\caption{\deltakid\ curves for \ion{D}{1}, \ion{O}{1} and
\ion{N}{1}. 
Each point on these curves corresponds to an individual fit, made 
with all the parameters free, but the \ion{D}{1}, \ion{O}{1}, or 
\ion{N}{1} column density fixed. \deltakid$=25$ corresponds to a 
$5\,\sigma$ error bar on the column density. \deltakid$=36$ corresponds to 
$6\,\sigma$. \deltakid$=49$ corresponds to $7\,\sigma$. 
\kid\ are rescaled on these plots (see Sect.~\ref{chi2_scaling}).
\label{fig_chi2_DON}
}
\end{figure}

For each estimated parameter, \eg\ $N$(\ion{D}{1}), error bars were
computed from an analysis of the \kid\ variation. 
We pegged that parameter at various trial
values, and for each trial value we ran an extra fit, allowing all the
other parameters to vary freely. 
We used these minima to define the confidence interval. 
Scanning the parameter in that
way for different values (up to at least \deltakid$\;=50$), we obtained
the value of \kid\ as a function of this parameter, and we derived the
1, 2, 3, 4, 5, 6, and $7\,\sigma$ error bars using the standard \deltakid\
method. The error bars at these different confidence levels have been taken 
into account to produce the final \ds\ uncertainties reported here.
The \deltakid\ curves as a function of column density are plotted in 
Figure~\ref{fig_chi2_DON}. 

However, since our best fit \kid\ value lies far above what would
be expected for the corresponding number of degrees of freedom, and since
that change is due to inaccurate estimates of the noise array, we rescaled
our \kid\ and the corresponding \deltakid\ values in the following way. At
the 95\% 
confidence level, the smallest value that one can draw from a
\kid\ distribution with 1764 degrees of freedom is 1667.5, and
consequently we increase the noise variance on each pixel by a factor
$2424.2/1667.5\simeq1.45$ (2424.2 corresponds to our best fit value), or
equivalently we divide our \kid\ and \deltakid\ by a factor 1.45. Note
that this is the most conservative choice at the 95\% 
confidence level for the
rescaling parameter. For example, 
for the $5\,\sigma$ contour in \kid\, one should
now look for differences \deltakid$\;=25\times1.45$ using our original error
array, which corresponds to the standard \deltakid$\;=25$ for the correctly
rescaled error array. This rescaling, similar to that presented 
by Lemoine et al.~(\citealp{lemoine01}), has increased in a 
conservative way the error bars derived~below.

\section{Results}
\label{Results}

\subsection{Parameters values}
\label{parameters_values}

Our column densities measurements are reported in 
Table~\ref{table_results}. 
We derived the following abundance ratios: 

$${\rm D}/{\rm O}  =  4.0\;(\pm1.2)\times 10^{-2}$$

$${\rm D}/{\rm N}  = 4.4\;(\pm1.3)\times 10^{-1}$$

\noindent
with \ds\ error bars.
These error bars were computed from the individual errors on 
the column densities listed in Table~\ref{table_results}, assuming 
that they are uncorrelated.
Although the noise may not follow a Gaussian distribution, 
we checked on smooth parts of the spectrum that the tabulated error bars 
scaled as described above (Sect.~\ref{chi2_scaling}) effectively 
include more than 95\% 
of the individual points of the spectrum. Therefore, the quoted ``\ds''
error bars can be considered as a 95\% 
confidence level, statistic and systematic effects being taking 
into account (see Sect.~\ref{systematic_effects}).

A formal temperature and turbulent velocity measured through the fit 
are $T=3400\pm2500$K and $\xi=3.0\pm1.0$\kms. 
These values should be taken with caution 
because several velocity components may be present and the spectral 
lines are not resolved. It is also a combination between 
turbulent velocity and velocity dispersion of the possible 
different clouds. In addition, 
these values were obtained from fits of unsaturated lines, which are 
not very sensitive to temperature and turbulent velocity. 

For a given channel, the relative shifts between different spectral 
windows were found to be relatively constant, with a dispersion 
below 10~\kms\ and a standard deviation of $\sim3$~\kms. 
Shifts between spectral windows of different channels were larger 
(up to 130~\kms). 

The full widths at half maximum obtained 
were $9.9\pm1.4$, $10.9\pm1.3$ and $10.9\pm1.2$ pixels respectively 
for spectral windows from HIRS, MDRS, and LWRS (values are averages 
and standard deviations on all the corresponding windows). As expected, 
the spectral resolution is slightly higher for HIRS than 
for MDRS and LWRS. All the widths found were in the range 8~--~13~pixels.
Note that for 14 spectral windows it was not possible to determine 
the LSF width because of the low signal-to-noise ratio; in 
these cases, the width was tuned to that obtained in the adjacent 
spectral windows.

\subsection{Systematic effects}
\label{systematic_effects}

\subsubsection{Overview}
\label{systematic_overview}

Error bars on column densities computed with the \deltakid\ method 
(Sect.~\ref{chi2_scaling}) 
include statistical {\it and} many possible systematic effects. 
The statistical aspect is obvious since the statistical error on each 
pixel is taken into account in the \kid\ computation. 

Several approaches were used in order to reduce systematic errors. 
Some effects were treated as free parameters (Sect.~\ref{free_parameters}).
Some systematic effects are likely to be decreased by 
fitting simultaneously several different lines for a given element. 
It is possible that some of these effects affect our fits but 
the use of numerous different lines for a given element 
(see, for example, Table~\ref{table_lines} for \ion{D}{1}, \ion{O}{1}, 
and \ion{N}{1}) averages out any possible bias. 
In the same spirit, effects which are totally or partially 
linked to instrumental effects are likely to be decreased by having 
data from multiple segments, multiple observations, and multiple slits 
fitted simultaneously. It is unlikely that one of these 
effects affect in the same manner a given line observed with what are 
essentially different instruments (different segments are used, 
through different slits, on different parts of the detectors). 
Finally, many possible systematic errors in the column density 
measurements are reduced by analyzing only unsaturated lines.

We list here the possible systematic effects which are treated using  
these four approaches:

\begin{itemize}

\item Free parameters: 
\begin{itemize}
\item continuum;
\item spectral shifts;
\item \lsf\ width.
\end{itemize}

\item Simultaneous fit of several lines for a given element: 
\begin{itemize}
\item line blending;
\item oscillator strengths uncertainty;
\item continuum;
\item zero flux level determination;
\item \fpn\ effects; 
\item \lsf\ effects;
\item flux calibration;
\item wavelength calibration. 
\end{itemize}

\item Simultaneous fit of several segments: 
\begin{itemize}
\item continuum;
\item zero flux level determination;
\item \fpn\ effects; 
\item \lsf\ effects;
\item flux calibration;
\item wavelength calibration. 
\end{itemize}

\item Unsaturated lines:  
\begin{itemize}
\item number of interstellar components;
\item temperature of the clouds;
\item turbulence of the clouds;
\item shape and width of the \lsf.
\end{itemize}

\end{itemize}

\subsubsection{Discussion of the effects and tests}
\label{discussion_of_the_effects_and_tests}

In this section we discuss all the effects we investigated and 
we present some tests that we carried out in order to check the 
reliability of the error bars.

\subsubsubsection{$\bullet$~Line blending} 

Line blending with unknown interstellar, stellar, or airglow features 
may affect the fit. However, such features are not present 
systematically for all lines of a given element. One exception 
is the \ion{He}{2} stellar lines, falling systematically near the 
interstellar \ion{D}{1} lines. 
However, no \ion{He}{2} absorption is seen in the \wdvd\ spectrum 
in the unblended lines at 1084.9\AA\ or~992.3\AA.

\subsubsubsection{$\bullet$~Oscillator strengths uncertainty}

Oscillator strengths ($f$) of \ion{D}{1} lines are well known. 
However, this is not the case for some lines 
of other elements, such as \ion{O}{1} or \ion{N}{1}. 
If some $f$-values were found to be inaccurate, the fit of that 
line would stand out as a poor fit and the line would be rejected 
(\eg, \ion{O}{1} $1026.47$\AA). 
By fitting each line individually, we checked 
that each tabulated $f$-value is consistent with the others 
(none of them present an abnormal tabulated $f$-value). 
The range in column densities from fits to the individual lines retained 
in the sample presented a $\sim20$\% 
dispersion around the value obtained with all the lines. 
Maybe some oscillator 
strengths are slightly inaccurate, but these errors should be small 
and average out in the final result. 

\subsubsubsection{$\bullet$~Continuum placement} 

The continua of the interstellar lines is given by the stellar continuum, 
possibly modified by instrumental effects (for example, one can see in 
Figure~\ref{fig_spectre_all} the bump near $1165$\AA\ on LiF1, which is 
due to an instrumental artifact). 
For each spectral window, the continuum fitted by a polynomial is 
near the actual continuum but is probably not a perfect model. 
However, when the continuum is fitted independently for each line, in 
spectra obtained in each segment and each spectrograph aperture, it is 
unlikely that the average placement will be systematically
higher or lower than the actual continuum.

We tested the possible systematic effect of the degree $n$ of 
the polynomials by performing fits with values of $n$ ranging from 
1 to 14. Column densities 
obtained were consistent within the \ds\ error bars.
Thus, there is no strong dependence of 
the fit on the degree of the polynomials used to fit the continua. 

Around the \ion{D}{1} lines, the continuum is affected by the 
blue wing of the \ion{H}{1} Lyman line. 
In order to check the \ion{D}{1} continuum levels found using polynomial 
approximations to the wings of the \ion{H}{1} Lyman lines, we performed 
additional fits with Voigt profiles used to describe the \ion{H}{1}
lines. The value of $N$(\ion{D}{1}) remained within the $2\sigma$ error
bar, so we infer that continuum levels at the \ion{D}{1} lines are not    
strongly dependent upon how the \ion{H}{1} lines are modeled. We note
that the \ion{H}{1} column density is not constrained well by these fits 
since the \ion{H}{1} Lyman lines lie on the flat part of the curve 
of growth. 

Finally, we divided the spectra by the stellar model we computed 
(Sect.~\ref{stellar_model}); fits of these normalized spectra 
produce column densities included within the \ds\ error~bars. Thus, 
no significant systematic effect linked to the continuum was found.

\subsubsubsection{$\bullet$~Zero flux level determination}

The zero flux level may be a function of wavelength. 
Simultaneous fits of different lines of a given element average out 
possible inaccuracies in the zero flux level determined from the bottom 
of the Lyman lines (Sect.~\ref{zero_flux_level}). 

We made extra fits in 
two extreme cases: doubling that correction (increasing the zero flux 
level by comparison with the best fit) and letting it remain zero. 
We also made fits where the zero flux level 
was a free parameter on each spectral window, with all windows fitted 
simultaneously. All these fits produced column density values again 
within the \ds\ error~bars.

\subsubsubsection{$\bullet$~Fixed-pattern noise effects}

Fixed-pattern noise 
is able to create false features that may artificially increase 
or decrease the intensity of an absorption line. 
The comparison of results for any single line, measured in different 
segments and apertures, allowed us to find lines affected by \fpn. 
In the final fit, no lines were found to be 
in strong disagreement with the other ones so these effects 
are small and average out in the fit.

\subsubsubsection{$\bullet$~Datasets}

In addition to fitting all the data simultaneously as described 
above, we also made fits to partial subsets of the total dataset
presented in Table~\ref{table_obs}. These subsets were chosen to find 
possible time-dependent effects, segment to segment, channel to channel 
or detector to detector differences, and slit (HIRS, MDRS, LWRS) effects 
such as optical transmission, \fpn, and airglow contamination 
(airglow intensity depends of the size of the aperture). 
All of the column densities found from these 
partial fits remained within the \ds\ error ranges, so no significant 
effects were found. As expected, error bars obtained with 
only one of these subsets of data were larger than the ones 
obtained with the complete dataset.

\subsubsubsection{$\bullet$~Reduction procedure}

Fits were made from observations reduced with 
different spectra extraction, 
different wavelength calibration files, or different extraction 
windows on the two-dimensional images of the detectors. 
We found no significant effects due to the calibration uncertainties.

\subsubsubsection{$\bullet$~Selection of lines}

In order to identify possible effects on a given line (such as 
\fpn\ right at the wavelength of a line or an erroneous 
$f$-value), we also fit each line independently of all 
others. No abnormal results were found for any line when compared 
to the composite fit results.
In addition, this allowed us to identify saturated lines to be 
excluded from the final fit. 
For example, fits to the \ion{O}{1} lines $\lambda \lambda948.69, 971.74$ 
and $1039.23$ (each line taken independently), 
with line strengths $f\lambda$ larger than $\sim5$\AA, 
had large uncertainties and overestimated the value of 
$N($\ion{O}{1}$)$ derived from weaker transitions.
We rejected the saturated lines using this objective~criterion.

\subsubsubsection{$\bullet$~Selection of the species} 

The \ion{D}{1}, \ion{O}{1}, \ion{N}{1}, \ion{Fe}{2}, and 
\ion{Si}{2} lines were fitted assuming that all these species were in 
the same interstellar (neutral) cloud. The same radial velocity, 
turbulence, and temperature were assumed for all these species. 
This assumption may be incorrect if the species do not co-exist
in the same regions along the sight line, which might be true if 
a significant amount of ionized gas is present. In order 
to test the possible systematic effect due to this assumption, 
we made extra fits including all these species but 
one. We also performed a fit including all the species but \ion{Fe}{2} 
and \ion{Si}{2}, which are the most likely ions to trace ionized gas. 
For all these fits, results for each column densities 
remained within the $1\,\sigma$ error bars.

\subsubsubsection{$\bullet$~Line spread function}

We used a simple Gaussian LSF in the fitting procedure and 
adopted a small zero flux level shift to account for weak, residual
light contributed probably by a broad component of the LSF 
(Sect.~\ref{zero_flux_level}). Note that 
the residual light does not affect the weak lines used in this
study in an appreciable way. The robustness of this approach was tested 
on a simulated spectra made with a double Gaussian LSF; they were fitted 
with a simple Gaussian LSF and a tuned zero flux level, and the result 
was satisfactory (see below). 

We also attempted to fit all spectral windows with a unique LSF width. 
Column densities obtained remained within the \ds\ 
error bars for LSF full widths at half maximum included in the range 
9~--~12.5~pixels; \deltakid\ of the 9 and 12.5~pixel LSF solutions were 
large ($>200$). 
Note that the minimum \kid\ was obtained for a 10.8~pixel width LSF.  
However, this \kid\ was greater by $\sim200$ than that obtained 
with the LSFs left free to vary from one spectral window to the next, 
showing that as expected, the LSF does depend on the wavelength, the 
segment, and the slit.

Finally, we performed fits with double Gaussian LSFs. We made tests using 
the LSF established by Wood et al.~(\citealp{wood01}), and other tests 
using free double Gaussian LSFs (Lemoine et al.~\citealp{lemoine01}). 
Fits obtained with double Gaussian LSFs gave the same column 
densities within the \ds\ error~bars.

\subsubsubsection{$\bullet$~Temperature and turbulent velocity}

The values obtained for $T$ and $\xi$ are inaccurate 
(Sect.~\ref{parameters_values}). We note, however, that the 
the impact of the uncertainty on the column densities is weak;
for temperatures and turbulent velocities within
0~--~20000K and 1.5~--~7.0\kms, the column densities are consistent 
within the \ds\ error bars. 
Hence, no effects on column densities due to $T$ or $\xi$ are 
significant, as expected from the fact that only unsaturated lines 
are fit.

\subsubsubsection{$\bullet$~Number of interstellar 
components on the sight line} 

Fits were made assuming only one interstellar component on the line 
of sight, even if it is likely that several components are present 
along this $\sim50$~pc sight line. Since we use only 
unsaturated lines, assuming only one interstellar component should 
have no effect on the measurement of the total column densities. 
Fits obtained with up to five interstellar components along the line of 
sight gave the same total column densities within the $1\,\sigma$ 
error~bars.

\subsubsubsection{$\bullet$~Simulation}

In order to test the various approaches of analyzing the interstellar 
lines from the \fuse\ spectra, an artificial spectrum was created for 
the \fuse\ Team members (Moos et al.~\citealp{moos01}). This simulated 
spectrum includes two interstellar clouds, \ion{He}{2} lines in the 
stellar continuum, random noise, \fpn, high frequency ringing, dithered 
wavelengths, background, and a double Gaussian LSF depending on 
the~segment. 
This simulated spectra was fit in a blind manner using the method 
presented above (including only one interstellar component and simple 
Gaussian LSF). All the estimated column densities were in agreement 
with the inputs in the simulation within the \ds\ error bars. 
This validates our approach and led us to a better understanding 
of the errors and the possible 
systematic uncertainties that might be present in our analysis. 

\subsubsubsection{$\bullet$~Method}

Finally, a subset of the data presented here has been independently 
fitted using the procedure presented by Wood et al.~(\citealp{wood01}). 
Both approaches gave similar results, in agreement within \ds. 
The final values listed in Table~\ref{table_results} reflect the 
combined effort of these~analyses.

\section{Discussion}
\label{Discussion}

The \ion{O}{1} and \ion{Ar}{1} column densities presented here 
agree with the first \fuse\ results reported by Jenkins et 
al.~(\citealp{jenkins00}). Agreement is poor for \ion{N}{1}, for 
which the value $14.02\pm0.15$ was reported by these authors. 
This early result was obtained with a lower number of lines and a lower 
signal-to-noise ratio. 
In the current study, the deficiency of \ion{N}{1} along this sight line 
with respect to the value in 
B stars appears not to be as large as reported by 
Jenkins et al.~(\citealp{jenkins00}). 

A value D/H$\;\simeq 1.5\times10^{-5}$ is obtained from 
the deuterium column density reported in Table~\ref{table_results} and  
the \ion{H}{1} column density obtained from EUVE spectra: 
$\log N($\ion{H}{1}$)=18.76$ (Wolff et al.~\citealp{wolff98}). 
No error bar is available for this model-dependent \ion{H}{1} estimate.
We note that the far-UV spectrum of \wdvd\ cannot be interpreted by using 
homogeneous H+He models (Marsh et al.~\citealp{marsh97b}), and that 
Holberg et al.~(\citealp{holberg98}) quoted a different result, 
$\log N($\ion{H}{1}$)=18.3$, based on fitting the \lya\ line 
observed with by IUE.

Another way to estimate D/H is to use our present D/O measurement 
together with the accurate abundance of interstellar oxygen obtained by 
Meyer et al.~(\citealp{meyer98}): O/H$\;=3.19 (\pm0.14) \times 10^{-4}$. 
The low dispersion of O/H confirms that \ion{O}{1} is a good 
tracer of \ion{H}{1} in the Galactic disk. 
The Meyer et al.~(\citealp{meyer98}) measurement was 
obtained from GHRS observations of 13 lines of sight ranging from 
130 to 1500~pc, with most of them closer than 500~pc. This study 
examines interstellar clouds that are more distant than \wdvd. 
However, we note that the 
O/H value given by Meyer et al.~(\citealp{meyer98}) 
agrees with that measured in the more local ISM 
(Moos et al.~\citealp{moos01}). 
Using the Meyer et al.~(\citealp{meyer98}) 
O/H value together with our new \fuse\ D/O measurement, we obtain 
D/H$\;=1.3 (\pm0.4) \times 10^{-5}$ (\ds). 
Note that using the revisited value O/H$\;=3.43 (\pm0.15) \times 10^{-4}$ 
from Meyer~(\citealp{meyer01}), which adopt an updated \ion{O}{1} 
oscillator strength, we obtain D/H$\;=1.4 (\pm0.4) \times 10^{-5}$ (\ds). 

In the same spirit, Meyer et al.~(\citealp{meyer97}) determined from 
GHRS observations toward 7 stars the interstellar ratio 
N/H$\;=7.5 (\pm0.4) \times 10^{-5}$. Adopting this value, our D/N 
value gives D/H$\;=3.3 (\pm1.2) \times 10^{-5}$ (\ds). 
We can probably understand the difference between the O-based and
N-based determinations of D/H in terms of the likelihood that N
could be somewhat more ionized than H, whereas O and H are more 
strongly coupled to each other by charge exchange reactions. 
For example, Vidal-Madjar et 
al.~(\citealp{avm98}) have found toward G191$-$B2B that the O/H is 
perfectly consistent with the average of Meyer et al.~(\citealp{meyer98}), 
while N/H is found to be a factor 2 smaller than the 
Meyer et al.~(\citealp{meyer97}) ratio. 
In the case of \wdvd, 
this possibility is supported by the fact that \ion{Ar}{1} 
is well below its expected column
density in relation to both \ion{N}{1} and \ion{O}{1}, using the abundance 
ratios of these elements in B-type stars as a reference (Cunha \& Lambert 
\citealp{cunha92},~\citealp{cunha94}; Holmgren et al.~\citealp{holmgren90}; 
Keenan et al.~\citealp{keenan90}).  According to arguments 
presented by Sofia \& Jenkins~(\citealp{sofia98}), this condition
indicates that a substantial fraction of the material is partially
ionized by EUV photons that are allowed to penetrate the region. 
At the same time, if $n_e\gg n_{\rm H\,{\sc I}}$ so that ordinary
recombinations with free electrons dominate over charge exchange
reactions of \ion{N}{2} with \ion{H}{1}, we might expect \ion{N}{1} 
to behave in a manner similar to \ion{Ar}{1}, 
\ie, $n_{\rm N\,{\sc II}}/n_{\rm N\,{\sc I}} > 
n_{\rm H\,{\sc II}}/n_{\rm H\,{\sc I}}$, 
since the neutral forms of both N and Ar have
photoionization cross sections larger than that of H.  This
inequality with the H ionization fraction is less likely to arise
with O, since the charge exchange cross section of \ion{O}{2} with 
\ion{H}{1} (Field~\& Steigman~\citealp{field71}) is considerably 
larger than for \ion{N}{2} with \ion{H}{1} (Butler 
\& Dalgarno~\citealp{butler79}).  An approximate indication of the
expected magnitudes of the deficiencies of \ion{N}{1} and \ion{Ar}{1} 
is shown in Figure~2 of Jenkins et al.~(\citealp{jenkins00}).

In any case, an accurate \ion{H}{1} column density value is needed 
to obtain a reliable \dshism\ measurement along this line of sight. 
Eventual observation of the Lyman~$\alpha$ toward \wdvd\ should 
provide this value since this line is not on the flat part of the 
curve of growth and therefore the damping wings can be used to measure 
$N$(\ion{H}{1}) more accurately. 
As of now, D/O and D/N estimates appear to be more robust than D/H for 
the \wdvd\ sight line. 

Since \ion{H}{1}, \ion{O}{1}, and \ion{D}{1} have 
nearly the same ionization potential, D/O ratio is an important 
proxy for the D/H ratio and its putative variations. 
D/O is very sensitive to astration, both from \ion{D}{1} 
destruction and \ion{O}{1} production (Timmes et al.~\citealp{timmes97}). 
If \fuse\ measurements of the D/O ratio toward other stars 
shows a relatively uniform value with low dispersion, it would 
argue in favor of D/H and O/H homogeneity in the \ism. Indeed, the only 
other possibility would be that D/H and O/H vary precisely in the same 
way as to cause D/O to remain constant. It seems improbable since (i) 
O/H appears to be uniform in the \ism\ over paths of several 
hundred parsecs, and (ii) astration should lead to 
an anti-correlation of \ion{D}{1} and \ion{O}{1} abundances. 
An upcoming survey of \ion{D}{1} and \ion{O}{1} absorption along about 
10 local lines of sight observed by \fuse\ should help determine whether 
the D/O ratio varies significantly from one sight line to the others 
(H\'ebrard et al.~\citealp{hebrard01}). Initial indications are that 
the ratio does not vary substantially (Moos et al.~\citealp{moos01}).

\section{Conclusion}
\label{Conclusion}

We have presented \fuse\ observations of the 905-1187~\AA\ spectrum of 
the white dwarf \wdvd. 
The column densities of several ions were measured through simultaneous 
fits to the numerous unsaturated absorption lines detected in the 
four channels and through the three apertures. 
In particular, the ratios D/O$\;= 4.0 \, (\pm1.2) \times 10^{-2}$ and 
D/N$\;= 4.4 \, (\pm1.3) \times 10^{-1}$ (\ds\ error bars) were 
measured. 
This result is discussed by Moos et al.~(\citealp{moos01}) together 
with other early \fuse\ deuterium results. 
Observations of the \lya\ absorption toward \wdvd\ would
help to constrain the D/H ratio along this sight line. 
Upcoming \fuse\ observations will help to determine if the D/H 
and D/O ratios vary locally. 
The answer to this question is critical to
improving our understanding of the abundance of deuterium.

\acknowledgments
This work is based on data obtained for the Guaranteed Time Team 
by the NASA-CNES-CSA \fuse\ mission operated by the Johns Hopkins 
University. Financial support to U. S. participants has been provided 
by NASA contract NAS5-32985. French participants are supported by CNES.



\begin{deluxetable}{lcc}
\tablecolumns{8}
\tablewidth{0pt}
\tablecaption{Target Summary for \wdvd}
\tablehead{
\colhead{Quantity} &
\colhead{Value} &
\colhead{Reference}
}
\startdata
 Spectral Type & DA & 1 \\
 $l$ & $345.79\degr$ & 2 \\
 $b$ & $-52.62\degr$ & 2 \\
 $d$ (pc)& 53 & 3 \\
 $V$ & 11.71 & 4 \\
$U-B$ & $-1.21$ & 4 \\
$B-V$ & $-0.34$ & 4\\
$T_{\rm eff}$ (K)& 62,000 & 5 \\
$\log g$ (cm s$^{-2}$)& 7.2 & 5 \\
$\log ({\rm He / H})$ & $<-4.0$ & 1 \\
$V_{\rm{rad}}$ (km s$^{-1}$) & +33.9 & 6 \\
\enddata
\label{wd_parameters}
\tablerefs{(1)\ Holberg et al.~(\citealp{holberg93}); 
(2)\ Hog et al.~(\citealp{hog98});
(3)\ Vennes et al.~(\citealp{vennes97}); 
(4)\ Marsh et al.~(\citealp{marsh97a}); 
(5)\ Barstow et al.~(\citealp{barstow98});
(6)\ Holberg et al.~(\citealp{holberg98}).}
\end{deluxetable}

\begin{deluxetable}{lcrrcc}
\tablecaption{Log of the observations.\label{table_obs}}
\tablewidth{0pt}
\tablehead{
\colhead{Obs. date\ \ \ } & 
\colhead{Obs. reference} & 
\colhead{$T_{\rm obs}$\tablenotemark{a}} & 
\colhead{$N_{\rm exp}$\tablenotemark{b}} & 
\colhead{Aperture\tablenotemark{c}} & 
\colhead{CALFUSE\tablenotemark{d}} 
}
\startdata
1999.10.25 &   M1030303  &  4.2      &   8  & LWRS  & 1.8.7 \\
1999.10.31 &   M1030304  &  3.0      &   6  & LWRS  & 1.8.7 \\
2000.06.03 &   M1030305  &  5.2      &  11  & LWRS  & 1.7.3 \\
2000.06.29 &   M1030306  &  4.2      &   7  & LWRS  & 1.7.7 \\
2000.08.17 &   M1030307  &  5.3      &  11  & LWRS  & 1.7.7 \\
2000.10.24 &   M1030308  &  4.1      &   9  & LWRS  & 1.7.7 \\
2000.10.24 &   M1030309  &  5.1      &  10  & LWRS  & 1.7.7 \\
2000.10.24 &   M1030310  &  5.8      &   8  & LWRS  & 1.7.7 \\
2000.10.25 &   M1030311  &  6.1      &  13  & LWRS  & 1.7.7 \\
2000.10.25 &   M1030312  &  5.5      &  10  & LWRS  & 1.7.7 \\
\hline
2000.06.03 &   P1043801  & 16.5      &  35  & MDRS  & 1.8.7 \\
2000.06.02 &   M1030201  &  4.8      &  10  & MDRS  & 1.8.7 \\
2000.08.18 &   M1030202  &  4.8      &  10  & MDRS  & 1.7.7 \\
2000.10.24 &   M1030203  &  4.8      &  10  & MDRS  & 1.7.7 \\
\hline
2000.06.02 &   M1030101  &  4.3      &   9  & HIRS  & 1.8.7 \\
2000.08.18 &   M1030102  &  5.8      &  12  & HIRS  & 1.8.7 \\
2000.10.24 &   M1030103  &  4.3      &   9  & HIRS  & 1.8.7 \\
\hline
\multicolumn{1}{r}{Total:\tablenotemark{e} } & 17 obs. & 93.8 & 188  \\
\enddata
\tablenotetext{a}{Total exposure time of the observation (in $10^3$~s).}
\tablenotetext{b}{Number of individual exposures during the observation.}
\tablenotetext{c}{LWRS, MDRS, and HIRS are respectively large, medium 
and narrow slits.}
\tablenotetext{d}{Version of the pipeline used for spectral extraction.}
\tablenotetext{e}{Totals for all datasets used in this study.}
\end{deluxetable}

\begin{deluxetable}{ccc|ccc|ccc}
\tablecaption{\ion{D}{1}, \ion{O}{1} and \ion{N}{1} lines included 
in the fit.\label{table_lines}}
\tablewidth{0pt}
\tablehead{
\colhead{$\lambda$\tablenotemark{a}} & \colhead{$f$\tablenotemark{b}} & 
\colhead{\#\tablenotemark{c}} &
\colhead{$\lambda$\tablenotemark{a}} & \colhead{$f$\tablenotemark{b}} & 
\colhead{\#\tablenotemark{c}} &
\colhead{$\lambda$\tablenotemark{a}} & \colhead{$f$\tablenotemark{b}} & 
\colhead{\#\tablenotemark{c}} 
}
\startdata
\multicolumn{3}{c|}{\ion{D}{1}} & \multicolumn{3}{|c|}{\ion{O}{1}} & 
\multicolumn{3}{|c}{\ion{N}{1}} \\
 & & & & & & & & \\
1025.4434 & $7.91\times10^{-2}$ & 9 & 
        930.2566  & $5.37\times10^{-4}$ & 5 &
                965.0413  & $4.02\times10^{-3}$ & 5 \\                
972.2723  & $2.90\times10^{-2}$ & 6 & 
        929.5168\tablenotemark{d}  & $2.36\times10^{-3}$ & 5 &
                964.6256  & $9.43\times10^{-3}$ & 6 \\
949.4847  & $1.39\times10^{-2}$ & 3 & 
        925.4420  & $3.54\times10^{-4}$ & 3 &
                963.9903  & $1.48\times10^{-2}$ & 3 \\
937.5484  & $7.80\times10^{-3}$ & 1 & 
        924.9520\tablenotemark{d}  & $1.59\times10^{-3}$ & 6 &
                954.1042  & $6.76\times10^{-3}$ & 6  \\
930.4951  & $4.82\times10^{-3}$ & 5 & 
        921.8570\tablenotemark{d}  & $1.19\times10^{-3}$ & 3 &
                953.9699  & $3.48\times10^{-2}$ & 6  \\
925.9737  & $3.18\times10^{-3}$ & 3 & 
        919.9080  & $1.78\times10^{-4}$ & 3  &
                953.6549  & $2.50\times10^{-2}$ & 6  \\
916.9311  & $7.23\times10^{-4}$ & 2 & 
        919.6580\tablenotemark{d}  & $9.47\times10^{-4}$ & 4  &
                953.4152  & $1.32\times10^{-2}$ & 6  \\
 & & & 
        916.8150\tablenotemark{d}  & $4.74\times10^{-4}$ & 2  &
                952.5227  & $6.00\times10^{-4}$ & 2  \\
 & & & 
         & & & 
                952.4148  & $1.70\times10^{-3}$ & 2  \\
 & & & 
         & & & 
                952.3034  & $1.87\times10^{-3}$ & 2  \\
\hline
\multicolumn{2}{r}{total:} & 29 & \multicolumn{2}{r}{total:} & 31 & 
\multicolumn{2}{r}{total:} & 44 \\
\enddata
\tablenotetext{a}{Wavelength in vacuum at rest (in \AA).}
\tablenotetext{b}{Oscillator strength.}
\tablenotetext{c}{Number of independent lines included in 
the fit (observed on different segments and through different 
slits).}
\tablenotetext{d}{Triplet structure used 
(see Morton~\citealp{morton91,morton99})}
\end{deluxetable}

\begin{deluxetable}{lc|lc}
\tablecaption{Total interstellar column densities. \label{table_results}}
\tablewidth{0pt}
\tablehead{
\colhead{species} & \colhead{$\log N$(\cmmd)\tablenotemark{a}} & 
\colhead{species} & \colhead{$\log N$(\cmmd)\tablenotemark{a}}
}
\startdata
\ion{D}{1} &  $13.94\;(\pm0.10)$  &
  \ion{Si}{2} &  $14.00\;(\pm0.15)$ \\
\ion{O}{1} &  $15.34\;(\pm0.08)$   &
  \ion{Ar}{1} &  $12.82\;(\pm0.10)$ \\
\ion{N}{1} &  $14.30\;(\pm0.06)$   &
  \ion{P}{2} &  $12.05\;(\pm0.20)$ \\
\ion{Fe}{2} &  $13.25\;(\pm0.15)$   &
  \ion{N}{2} &  $14.6\;(\pm0.3)$ \\
\enddata
\tablenotetext{a}{$2\sigma$ error~bars.}
\end{deluxetable}


\begin{thebibliography}{}
\bibitem[1992]{allen92} 
  Allen, M.\ M., Jenkins, E.\ B., \& Snow, T.\ P.\ 1992, \apjs, 83, 261 
\bibitem[1998]{barstow98} 
  Barstow, M.\ A., Hubeny, I., \& Holberg, J.\ B.\ 1998, \mnras, 299, 520 
\bibitem[1987]{blitz87} 
  Blitz, L.\ \& Heiles, C.\ 1987, \apj, 313, L95 
\bibitem[1998a]{burles98a} 
  Burles, S.\ \& Tytler, D.\ 1998a, \apj, 499, 699 
\bibitem[1998b]{burles98b} 
  Burles, S.\ \& Tytler, D.\ 1998b, \apj, 507, 732 
\bibitem[1979]{butler79}
  Butler, S. E., \& Dalgarno, A. 1979, \apj, 234, 765
\bibitem[2001]{chayer01} 
  Chayer, P., et al. 2001, in preparation
\bibitem[1997]{chenga97} 
  Chengalur, J.\ N., Braun, R., \& Burton, W.\ B. \ 1997, \aap, 318, L35 
\bibitem[1992]{cunha92} 
  Cunha, K., \& Lambert, D. L. 1992, \apj, 399, 586
\bibitem[1994]{cunha94} 
  Cunha, K., \& Lambert, D. L. 1994, ApJ, 426, 170
\bibitem[2000]{bernardis00} 
  de Bernardis, P.\ et al.\ 2000, \nat, 404, 955 
\bibitem[2001]{bernardis01} 
  de Bernardis, P.\ et al.\ 2001, \apj, submitted ({\tt astro-ph/0105296})
\bibitem[1976]{epstein76} 
  Epstein, R.\ I., Lattimer, J.\ M., \& Schramm, D.\ N. 
      1976, \nat, 263, 198
\bibitem[2000]{ferlet00} 
  Ferlet, R. et al. 2000, \apj, 438, L69
\bibitem[1971]{field71} 
  Field, G. B., \& Steigman, G. 1971, \apj, 166, 59
\bibitem[1997]{finley97} 
  Finley, D.\ S., Koester, D., \& Basri, G.\ 1997, \apj, 488, 375 
\bibitem[2001]{friedman01} 
  Friedman, S. D. et al. 2001, accepted for publication in \apjs\ 
       ({\tt astro-ph/0111332})
\bibitem[1995]{galli95} 
  Galli, D., Palla, F., Ferrini, F., \& Penco, U. 1995, \apj, 443, 536
\bibitem[1999]{hebrard99} 
  H\'ebrard, G., Mallouris, C., Ferlet, R., Koester, D., Lemoine, M., 
      Vidal-Madjar, A., \& York, D.\ 1999, \aap, 350, 643 
\bibitem[2000]{hebrard00} 
  H\'ebrard, G., P{\'e}quignot, D., Walsh, J.\ R., Vidal-Madjar, A., 
      \& Ferlet, R.\ 2000, \aap, 364, L31 
\bibitem[2001]{hebrard01} 
  H\'ebrard, G. et al. 2001, XVIIth IAP Colloquium, {\it Gaseous 
matter in galaxies and intergalactic space}, 19-23 June 2001, Paris, 
Edited by R. Ferlet et al., to be published; \apj, in preparation
\bibitem[1998]{hog98} 
  Hog, E., Kuzmin, A., Bastian, U., Fabricius, C., Kuimov, K., 
      Lindegren, L., Makarov, V.\ V., \& Roeser, S.\ 1998, \aap, 335, 65
\bibitem[1993]{holberg93} 
  Holberg, J.\ B.\ et al.\ 1993, \apj, 416, 806 
\bibitem[1994]{holberg94} 
  Holberg, J.\ B., Hubeny, I., Barstow, M.\ A., Lanz, T., Sion, E.\ M.,
       \& Tweedy, R.\ W.\ 1994, \apj, 425, L105
\bibitem[1998]{holberg98} 
  Holberg, J.\ B., Barstow, M.\ A., \& Sion, E.\ M.\ 1998, \apjs, 119, 207 
\bibitem[1990]{holmgren90} 
  Holmgren, D. E., Brown, P. J. F., Dufton, P. L., \& Keenan, F. P. 1990, 
       \apj, 364, 657
\bibitem[1995]{hubeny95} 
  Hubeny, I.\ \& Lanz, T.\ 1995, \apj, 439, 875 
\bibitem[2001]{jaffe01} 
  Jaffe, A. H., et al. 2001, \prl, 86, 3475
\bibitem[1999]{jenkins99} 
  Jenkins, E.\ B., Tripp, T.\ M., Wo{\'z}niak, P.\ ;.\ A., Sofia, U.\ J., 
      \& Sonneborn, G.\ 1999, \apj, 520, 182 
\bibitem[2000]{jenkins00} 
  Jenkins, E.\ B.\ et al.\ 2000, \apjl, 538, L81 
\bibitem[1990]{keenan90} 
  Keenan, F. P., Bates, B., Dufton, P. L., Holmgren, D. E., \& Gilheany, S. 
      1990, \apj, 348, 322  
\bibitem[2001]{kruk01} 
  Kruk, J. W. et al. 2001, accepted for publication in \apjs
\bibitem[1992]{lallement92} 
  Lallement, R., \& Bertin, P. 1992, \aap, 266, 479
\bibitem[1979]{laurent79} 
  Laurent, C., Vidal-Madjar, A., \& York, D.\ G.\ 1979, \apj, 229, 923 
\bibitem[2001]{lehner01} 
  Lehner, N. et al. 2001, \apjs, accepted for publication in \apjs\
        ({\tt astro-ph/0111336})
\bibitem[1996]{lemoine96} 
  Lemoine, M., Vidal-Madjar, A., Bertin, P., Ferlet, R., Gry, C., 
      \& Lallement, R.\ 1996, \aap, 308, 601 
\bibitem[1999]{lemoine99} 
  Lemoine, M.\ et al.\ 1999, New Astronomy, 4, 231 
\bibitem[2001]{lemoine01} 
  Lemoine, M. et al. 2001, accepted for publication in \apjs\ 
     ({\tt astro-ph})
\bibitem[1998]{linsky98} 
  Linsky, J.\ L.\ 1998, Space Science Reviews, 84, 285 
\bibitem[2000]{lubo00} 
  Lubowich, D.\ A., Pasachoff, J.\ M., Balonek, T.\ J., Millar, 
      T.\ J., Tremonti, C., Roberts, H., \& Galloway, R.\ P.\ 2000, 
      \nat, 405, 1025 
\bibitem[1997a]{marsh97a} 
  Marsh, M.\ C.\ et al.\ 1997a, \mnras, 286, 369 
\bibitem[1997b]{marsh97b}
  Marsh, M.\ C.\ et al.\ 1997b, \mnras, 287, 705 
\bibitem[1997]{meyer97} 
  Meyer, D.\ M., Cardelli, J.\ A., \& Sofia, U.\ J.\ 1997, \apjl, 
      490, L103 
\bibitem[1998]{meyer98} 
  Meyer, D.\ M., Jura, M., \& Cardelli, J.\ A.\ 1998, \apj, 493, 222 
\bibitem[2001]{meyer01} 
  Meyer, D.\ M., XVIIth IAP Colloquium, {\it Gaseous 
      matter in galaxies and intergalactic space}, 19-23 June 2001, Paris, 
      Edited by R. Ferlet et al., to be published
\bibitem[2000]{moos00} 
  Moos, H.\ W. et al. 2000, \apj, 538, L1
\bibitem[2001]{moos01} 
  Moos, H.\ W., Sembach, K. R. et al. 2001, accepted for publication in \apjs
\bibitem[1991]{morton91} 
  Morton, D.\ C.\ 1991, \apjs, 77, 119 
\bibitem[1999]{morton99} 
  Morton, D.\ C.\ 1999, private communication 
    ({\tt http://www.hia.nrc.ca/staff/dcm/atomic\_data.html})
\bibitem[1993]{pounds93}
  Pounds, K.\ A.\ 1993, \mnras, 260, 77
\bibitem[1996]{prantzos96} 
  Prantzos, N. 1996, \aap, 310, 106
\bibitem[2001]{pryke01} 
  Pryke, C.\ et al.\ 2001, \apj, submitted ({\tt astro-ph/0104490})
\bibitem[1973]{ry73} 
  Rogerson, J., \& York, D.\ 1973, \apj, 186, L95
\bibitem[2000]{sahnow00} 
  Sahnow, D.\ J.\ et al.\ 2000, \apjl, 538, L7 
\bibitem[1999]{sahu99} 
  Sahu, M.\ S.\ et al.\ 1999, \apjl, 523, L159 
\bibitem[1997]{scully97} 
  Scully, S. T., Cass\'e, M., Olive. K. A., \& Vangioni-Flam, E. 1997, 
      \apj, 476, 521
\bibitem[1998]{sofia98}
  Sofia, U. J., \& Jenkins, E. B. 1998, \apj, 499, 951
\bibitem[2000]{sonneborn00} 
  Sonneborn, G., Tripp, T.\ M., Ferlet, R., Jenkins, E.\ B., Sofia, 
      U.\ J., Vidal-Madjar, A., \& Wo{\'z}niak, P.\ ;.\ R.\ 2000, 
      \apj, 545, 277 
\bibitem[2001]{sonneborn01} 
  Sonneborn, G. et al. 2001, \apjs, submitted
\bibitem[2001]{stompor01} 
  Stompor, R. et al. 2001, \apj, submitted ({\tt astro-ph/0105062})
\bibitem[1997]{timmes97} 
  Timmes, F.\ X., Truran, J.\ W., Lauroesch, J.\ T., \& York, D.\ G.\ 
      1997, \apj, 476, 464 
\bibitem[1998]{tosi98} 
  Tosi, M., Steigman, G., Matteucci, F., \& Chiappini, C.\ 1998, 
      \apj, 498, 226
\bibitem[1994]{flam94} 
  Vangioni-Flam, E., Olive, K. A., et al. 1994, \apj, 427, 618
\bibitem[1995]{flam95} 
  Vangioni-Flam, E., \& Cass\'e, M. 1995, \apj, 441, 471
\bibitem[1997]{vennes97} 
  Vennes, S., Thejll, P.\ A., Galvan, R.\ G., \& Dupuis, J.\ 1997,
      \apj, 480, 714
\bibitem[2000]{vennes00} 
  Vennes, S., Polomski, E.\ F., Lanz, T., Thorstensen, J.\ R., 
      Chayer, P., \& Gull, T.\ R.\ 2000, \apj, 544, 423 
\bibitem[1998]{avm98} 
  Vidal-Madjar, A.\ et al.\ 1998, \aap, 338, 694 
\bibitem[1997]{webb97} 
  Webb, J.\ K., Carswell, R.\ F., Lanzetta, K.\ M., Ferlet, R., Lemoine, 
      M., Vidal-Madjar, A., \& Bowen, D.\ V.\ 1997, \nat, 388, 250 
\bibitem[1994]{werner94} 
  Werner, K., \& Dreizler, S.\ 1994, \aap, 286, L31
\bibitem[1998]{wolff98} 
  Wolff, B., Koester, D., Dreizler, S., \& Haas, S.\ 1998, 
      \aap, 329, 1045 
\bibitem[2001]{wood01} 
  Wood, B.\ E. et al. 2001, accepted for publication in \apjs
\bibitem[1983]{york83} 
  York, D.\ G.\ 1983, \apj, 264, 172 
\end{thebibliography}
\end{document}